\documentclass[dvips]{acta}
\usepackage{supertabular,lscape,epsfig}
\usepackage{amssymb}
\usepackage{amsmath}
\usepackage{multirow}
\usepackage{float}
\mathchardef\mhyphen="2D
\usepackage{placeins}
\DeclareSymbolFont{ppa}{OT1}{ppl}{m}{it}
\DeclareMathSymbol{\vv}{\mathalpha}{ppa}{'166}

\SetPages{189}{204}

\SetVol{71}{2021}

\begin{document}
\newcommand\pvalue{\mathop{p\mhyphen {\rm value}}}
\newcommand{\TabApp}[2]{\begin{center}\parbox[t]{#1}{\centerline{
  {\bf Appendix}}
  \vskip2mm
  \centerline{\small {\spaceskip 2pt plus 1pt minus 1pt T a b l e}
  \refstepcounter{table}\thetable}
  \vskip2mm
  \centerline{\footnotesize #2}}
  \vskip3mm
\end{center}}

\newcommand{\TabCapp}[2]{\begin{center}\parbox[t]{#1}{\centerline{
  \small {\spaceskip 2pt plus 1pt minus 1pt T a b l e}
  \refstepcounter{table}\thetable}
  \vskip2mm
  \centerline{\footnotesize #2}}
  \vskip3mm
\end{center}}

\newcommand{\TTabCap}[3]{\begin{center}\parbox[t]{#1}{\centerline{
  \small {\spaceskip 2pt plus 1pt minus 1pt T a b l e}
  \refstepcounter{table}\thetable}
  \vskip2mm
  \centerline{\footnotesize #2}
  \centerline{\footnotesize #3}}
  \vskip1mm
\end{center}}

\newcommand{\MakeTableH}[4]{\begin{table}[H]\TabCap{#2}{#3}
  \begin{center} \TableFont \begin{tabular}{#1} #4 
  \end{tabular}\end{center}\end{table}}

\newcommand{\MakeTableApp}[4]{\begin{table}[p]\TabApp{#2}{#3}
  \begin{center} \TableFont \begin{tabular}{#1} #4 
  \end{tabular}\end{center}\end{table}}

\newcommand{\MakeTableSepp}[4]{\begin{table}[p]\TabCapp{#2}{#3}
  \begin{center} \TableFont \begin{tabular}{#1} #4 
  \end{tabular}\end{center}\end{table}}

\newcommand{\MakeTableee}[4]{\begin{table}[htb]\TabCapp{#2}{#3}
  \begin{center} \TableFont \begin{tabular}{#1} #4
  \end{tabular}\end{center}\end{table}}

\newcommand{\MakeTablee}[5]{\begin{table}[htb]\TTabCap{#2}{#3}{#4}
  \begin{center} \TableFont \begin{tabular}{#1} #5 
  \end{tabular}\end{center}\end{table}}


\newcommand{\MakeTableHH}[4]{\begin{table}[H]\TabCapp{#2}{#3}
  \begin{center} \TableFont \begin{tabular}{#1} #4 
  \end{tabular}\end{center}\end{table}}

\newfont{\bb}{ptmbi8t at 12pt}
\newfont{\bbb}{cmbxti10}
\newfont{\bbbb}{cmbxti10 at 9pt}
\newcommand{\uprule}{\rule{0pt}{2.5ex}}
\newcommand{\douprule}{\rule[-2ex]{0pt}{4.5ex}}
\newcommand{\dorule}{\rule[-2ex]{0pt}{2ex}}
\def\thefootnote{\fnsymbol{footnote}}
\begin{Titlepage}
\Title{Over 24\,000 $\delta$~Scuti Stars in the Galactic Bulge and Disk\\
from the OGLE Survey\footnote{Based on observations obtained with the
1.3-m Warsaw telescope at the Las Campanas Observatory of the Carnegie
Institution for Science.}}
\vspace*{-3pt}
\Author{I.~~S~o~s~z~y~\'n~s~k~i$^1$,~~
P.~~P~i~e~t~r~u~k~o~w~i~c~z$^1$,~~
J.~~S~k~o~w~r~o~n$^1$,~~
A.~~U~d~a~l~s~k~i$^1$,\\
M.\,K.~~S~z~y~m~a~\'n~s~k~i$^1$,~~
D.\,M.~~S~k~o~w~r~o~n$^1$,~~
R.~~P~o~l~e~s~k~i$^1$,~~
S.~~K~o~z~\l~o~w~s~k~i$^1$,\\
P.~~M~r~\'o~z$^1$,~~
K.~~U~l~a~c~z~y~k$^{2,1}$,~~
K.~~R~y~b~i~c~k~i$^1$,~~
P.~~I~w~a~n~e~k$^1$,~~
M.~~W~r~o~n~a$^1$~~\\
and~~M.~~G~r~o~m~a~d~z~k~i$^1$
}
{$^1$Astronomical Observatory, University of Warsaw, Al.~Ujazdowskie~4,\\ 00-478~Warszawa, Poland\\
$^2$Department of Physics, University of Warwick, Gibbet Hill Road, Coventry, CV4~7AL,~UK}
\Received{November 3, 2021}
\end{Titlepage}

\Abstract{We present the largest collection of
  $\delta$~Scuti-type stars in the Milky Way. Together with the
  recently published OGLE collection of $\delta$~Sct variables in the
  inner Galactic bulge, our sample consists of 24\,488 objects
  distributed along the Milky Way plane, over galactic longitudes
  ranging from about $-170\arcd$ to $+60\arcd$. The collection data
  include the {\it I}- and {\it V}-band time-series photometry
  collected since 1997 during the OGLE-II, OGLE-III, and OGLE-IV
  surveys. We show the on-sky distribution of $\delta$~Sct stars in
  the Galactic bulge and disk, discuss their period, luminosity and
  amplitude distributions, present Petersen diagram for multimode
  pulsators, distinguish 34~$\delta$~Sct stars in eclipsing and
  ellipsoidal binary systems, and list probable members of globular
  clusters.}{Stars: variables: delta Scuti -- Stars: oscillations --
  Galaxy: bulge -- Galaxy: disk -- Catalogs}

\Section{Introduction}
$\delta$~Sct variables constitute a diverse class of short-period
pulsating stars located in the lower part of the classical instability
strip in the Hertzsprung-Russell diagram. These objects cover a wide range
of stellar evolutionary stages: young stellar objects during their
contraction toward the main sequence, stars with core hydrogen burning on
the main sequence, subgiants evolving off the main sequence, and blue
stragglers known as SX~Phoenicis variables\footnote{In this paper, we refer
to SX~Phe stars as one of the subgroups of $\delta$~Sct stars.}.
$\delta$~Sct pulsators can be monoperiodic or multiperiodic, with radial
or nonradial modes excited, and the pulsation periods ranging from
approximately 0.03~d to 0.3~d. Like other types of classical pulsators
(Cepheids and RR~Lyrae stars), $\delta$~Sct variables follow
period--luminosity relations, which makes them potentially useful as
distance indicators.

The number of known $\delta$~Sct stars in the Milky Way has grown
rapidly in recent years, mostly thanks to the large-scale photometric
sky surveys. The catalog of all $\delta$~Sct variables known in
January 2000 compiled by Rodr{\'\i}guez \etal (2000) consisted of 636
objects. This number has increased by a factor of $\approx50$ since
then. So far, the largest samples of the Galactic $\delta$~Sct stars
have been included in the ASAS-SN catalogue of variable stars
($\approx8400$ objects, Jayasinghe \etal 2020) and in the Zwicky
Transient Facility Catalog of Periodic Variable Stars
($\approx15\,000$ objects, Chen \etal 2020).

Recently, Pietrukowicz \etal (2020) published a collection of 10\,092
$\delta$~Sct stars toward the Galactic bulge detected in the Optical
Gravitational Lensing Experiment (OGLE) photometric database. About
97\% of this sample were new discoveries. In this paper, we present an
extension of the Pietrukowicz \etal (2020) catalog to include newly
detected $\delta$~Sct stars in the Galactic disk and outer bulge --
regions of the sky photometrically monitored by the OGLE Galaxy
Variability Survey (GVS). Our extended collection contains over
$24\,000$ $\delta$~Sct variables and is currently the largest sample
of such objects in the Milky Way. This is also a part of the OGLE
Collection of Variable Stars (OCVS) consisting of over a million
carefully selected and classified variable stars of various
types. Among others, we recently published extensive catalogs of
Cepheids (Udalski \etal 2018, Soszy\'nski \etal 2020) and RR~Lyr stars
(Soszy\'nski \etal 2019) in the Milky Way.

The outline of the paper is as follows. In the next section, we
present the OGLE observations and data reduction pipeline. Section~3
provides a description of the procedures used for the identification
and classification of variable stars in the OGLE photometric
database. In Section~4, we summarize the OGLE collection of
$\delta$~Sct stars in the Galactic bulge and disk and present their
on-sky distribution. In Section~5, we cross-match our sample to
external catalogs and estimate its completeness. Particularly
interesting cases of $\delta$~Sct variables: multimode pulsators,
binary systems containing pulsating components, members of globular
clusters, are discussed in Section~6. Finally, Section~7 summarizes
our results.

\vspace*{9pt}
\Section{Observations and Data Reduction}
\vspace*{5pt}
The OGLE photometric observations used in this work were carried out
with the 1.3-m Warsaw telescope at Las Campanas Observatory, Chile,
between 2013 and 2020. The field of view of the OGLE CCD mosaic camera
is 1.4~square degrees and the pixel scale is 0\zdot\arcs26. The
telescope is equipped with two filters, {\it I}- and {\it V}-bands,
closely resembling Cousins-Johnson standard photometric
systems. Details of the OGLE instrumentation and data reduction
procedures can be found in Udalski \etal (2015a, 2018).

The OGLE GVS project covers approximately 3000 square degrees along
the Galactic plane, from galactic longitudes of about $-170\arcd$ to
$+60\arcd$ and latitudes from $-7\arcd$ to $+7\arcd$ in the disk and
from $-15\arcd$ to $+15\arcd$ in the bulge region. This is a shallower
survey than the regular OGLE project observations, with 25~s and 30~s
integration times for {\it I}-band and {\it V}-band, respectively
(compared to 100--150~s for the regular OGLE observations).

Until March 2020, when the Warsaw telescope was temporarily closed due
to the COVID-19 pandemic, between 100 and 200 {\it I}-band epochs per
star have been acquired for the majority of the GVS fields. Time span
of individual light curves range from 2~yr to 7~yr. The {\it V}-band
observations (usually several epochs) are currently available for
approximately half of the stars.

\Section{Selection and Classification of $\delta$~Sct Stars}
Our search for Galactic $\delta$~Sct stars was very similar to the
search for RR Lyr stars and Cepheids in the same fields (Udalski \etal
2018, Soszy\'nski \etal 2019, 2020). In the first step, over 1~billion
point sources in the outer bulge and Galactic disk photometrically
monitored by the OGLE GVS were subjected to a Fourier-based frequency
analysis with the {\sc Fnpeaks} code \footnote{\it
  http://helas.astro.uni.wroc.pl/deliverables.php?lang=en\&active=fnpeaks}. The
probed frequency space ranged from 0 to 30 cycles per day, with a step
of $5\cdot10^{-5}$ cycles per day. For each analyzed time series, the
{\sc Fnpeaks} code returned 10 most significant periodicities with
their signal-to-noise ratios. Then, each light curve was prewhitened
with the primary frequency and its harmonics and the period search was
repeated on the residuals.

In the second step, we performed a visual inspection of {\it I}-band
light curves with the primary periods below 0.3~d and the highest
signal-to-noise ratios. In this way, we divided our targets into three
groups: candidates for pulsating stars, eclipsing and ellipsoidal
binaries, and other (usually unknown) types of variable stars. Our
classification criteria were mainly based on the morphology of the
light curves, in particular we required an asymmetry seen in the
phased light curve to consider a star as a pulsator. However, the
above condition was not applied to multiperiodic pulsating variables
-- in these cases we primarily relied on their characteristic period
ratios (see Section~6.2) and their position in the Petersen diagram
(where the shorter-to-longer period ratio is plotted against the
logarithm of the longer period).

In the last step of our procedure, we isolated probable $\delta$~Sct
variables from the list of candidate pulsating stars. Since there is a
continuity between $\delta$~Sct stars and classical Cepheids, we
adopted a period of 0.23~d of the first-overtone mode (corresponding
to about 0.3~d for the fundamental mode) to separate both classes of
pulsators. The distinction between long-period ($P>0.2$~d)
$\delta$~Sct stars and short-period ($P<0.3$~d) overtone RR~Lyr stars
is not obvious. Most of the single-mode pulsating stars with periods
in this range were recognized as RRc variables and published in the
OGLE collection of RR Lyr stars (Soszy\'nski \etal 2019). On the other
hand, multimode variables with these periods were mostly classified as
$\delta$~Sct stars, because double-mode RR~Lyr (RRd) stars generally
have longer periods and different period-ratios. Finally, we carefully
checked and rejected most of the stars with $(V-I)$ color index below
0.2~mag, since such blue objects probably belong to the $\beta$~Cephei
class of pulsating variables. However, it cannot be ruled out that a
number of $\beta$~Cep stars with the colors reddened by a strong
interstellar extinction remained in our list.  Spectroscopic
observations are required to ultimately distinguish $\beta$~Cep from
$\delta$~Sct pulsators.

Similarly, the division of $\delta$~Sct variables into Population I
and Population II (SX Phe) stars requires complex studies of their
evolutionary status and is beyond the scope of this paper. The
distinction between $\delta$~Sct and SX~Phe stars is not possible on
the basis of their amplitudes or number of frequencies (Balona and
Nemec 2012), also there is no boundary period that can be used to
separate both populations, although SX Phe stars have on average
shorter periods than their younger counterparts (Jayasinghe \etal
2020). There are also difficulties in unambiguously identifying the
pulsation modes in some individual $\delta$~Sct stars, so we decided
not to provide mode classification in our collection. Instead, we
split our sample into singlemode and multimode variables. The latter
group contains objects with well-defined secondary periods that can be
associated with additional radial or nonradial pulsation modes.

In total, we distinguished 14\,328 likely Galactic $\delta$~Sct stars
that were not included in the previously published parts of the OCVS
(Pietrukowicz \etal 2013, 2020). This sample has enlarged our
collection of $\delta$~Sct variables in the Milky Way by 140\%, and
currently this is the largest catalog of such stars.

\Section{The OGLE Collection of Galactic $\delta$~Sct Stars}

The newly detected $\delta$~Sct stars in the Galactic bulge and disk have
been added to the already published parts of the OCVS (Pietrukowicz \etal
2013, 2015, 2020). Currently, the full collection contains 24\,488
$\delta$~Sct stars divided into 18\,638 single-mode and 5850 multimode
pulsators. 16\,812 of these stars have been found in the bulge region,
while 7676 objects are located in the Galactic disk\footnote{The bulge and
disk OGLE fields are shown in the maps on the OGLE website.}. All sources
are available online at the OGLE Internet Archive:
\begin{center}
{\it https://ogle.astrouw.edu.pl $\rightarrow$ OGLE Collection of Variable Stars}\\
{\it https://www.astrouw.edu.pl/ogle/ogle4/OCVS/blg/dsct/}\\
{\it https://www.astrouw.edu.pl/ogle/ogle4/OCVS/gd/dsct/}\\
\end{center}

The identifiers of our $\delta$~Sct stars follow the scheme introduced
by Pietrukowicz \etal (2013, 2020): OGLE-BLG-DSCT-XXXXX and
OGLE-GD-DSCT-YYYY for bulge and disk populations, respectively, where
XXXXX and YYYY are consecutive numbers. The new samples are arranged
according to their equinox equatorial coordinates ordered by
increasing right ascension. For each star, we provide its equatorial
coordinates (J2000.0), intensity-averaged mean magnitudes in the {\it
  I}- and {\it V}-bands (if available), and up to three dominant
periods with the corresponding amplitudes, epochs of the maximum
light, and Fourier coefficients. The pulsation periods have been
refined with the {\sc Tatry} code (Schwarzenberg-Czerny 1996). For the
previously known variables, we give their designations from the
International Variable Star Index (VSX, Watson \etal 2006).

\vskip4pt The OGLE {\it I}- and {\it V}-band time-series photometry
collected during the OGLE-II, OGLE-III, and OGLE-IV projects (if
available) have been provided for the entire sample, so the time span
of the light curves may reach 23~yr (1997--2020) for some $\delta$~Sct
stars. The photometric data from each phase of the OGLE project have
been independently calibrated to the standard Johnson-Cousins
photometric system. However, please note that small offsets between
the photometric zero points may occur for individual objects, which
can be a result of crowding and blending by unresolved stars or
different instrumental configurations during the three phases of the
OGLE survey. These zero-point shifts should be compensated before
merging the light curves from different stages of the project.

\vskip4pt
Fig.~1 shows on-sky maps of our collection of $\delta$~Sct stars. The
upper panel presents a surface density map in a square root
scaling. The best-populated fields in the Galactic bulge contain over
200 $\delta$~Sct variables per square degree, while around the
Galactic anticenter we found on average only 2.7 $\delta$~Sct stars
per square degree. Additionally, there is a deficit of variables close
to the Galactic plane in the bulge region, which is due to the
enormous interstellar extinction in these directions.

\vskip4pt
The middle and lower panels of Fig.~1 show positions of individual
$\delta$~Sct stars plotted over the OGLE-IV footprint. Different
colors of the points refer to different mean {\it I}-band magnitudes
and different dominant pulsation periods for middle and lower panels,
respectively. Both distributions are not homogeneous. The distribution
of magnitudes (middle panel of Fig.~1) primarily shows the map of
interstellar matter close to the Galactic plane, while the period
distribution (lower panel of Fig.~1) primarily reflects the
metallicity gradient in the Milky Way.

\vskip4pt
As can be seen by inspection of the lower panel of Fig.~1, stars
located farther from the Galactic equator have generally shorter
periods, which is especially noticeable in the bulge
region. Jayasinghe \etal (2020) noticed that metal-poor $\delta$~Sct
stars have on average shorter pulsation periods than metal-rich
variables, so the visible period gradient can be useful for
distinguishing between Population~II SX~Phe stars (members of the
spheroidal component of the Milky Way) and Population~I $\delta$~Sct
variables (disk members).

\begin{landscape}
\begin{figure}[p]
\centerline{\includegraphics[bb = 80 20 495 800, clip, width=11.3cm, angle=-90]{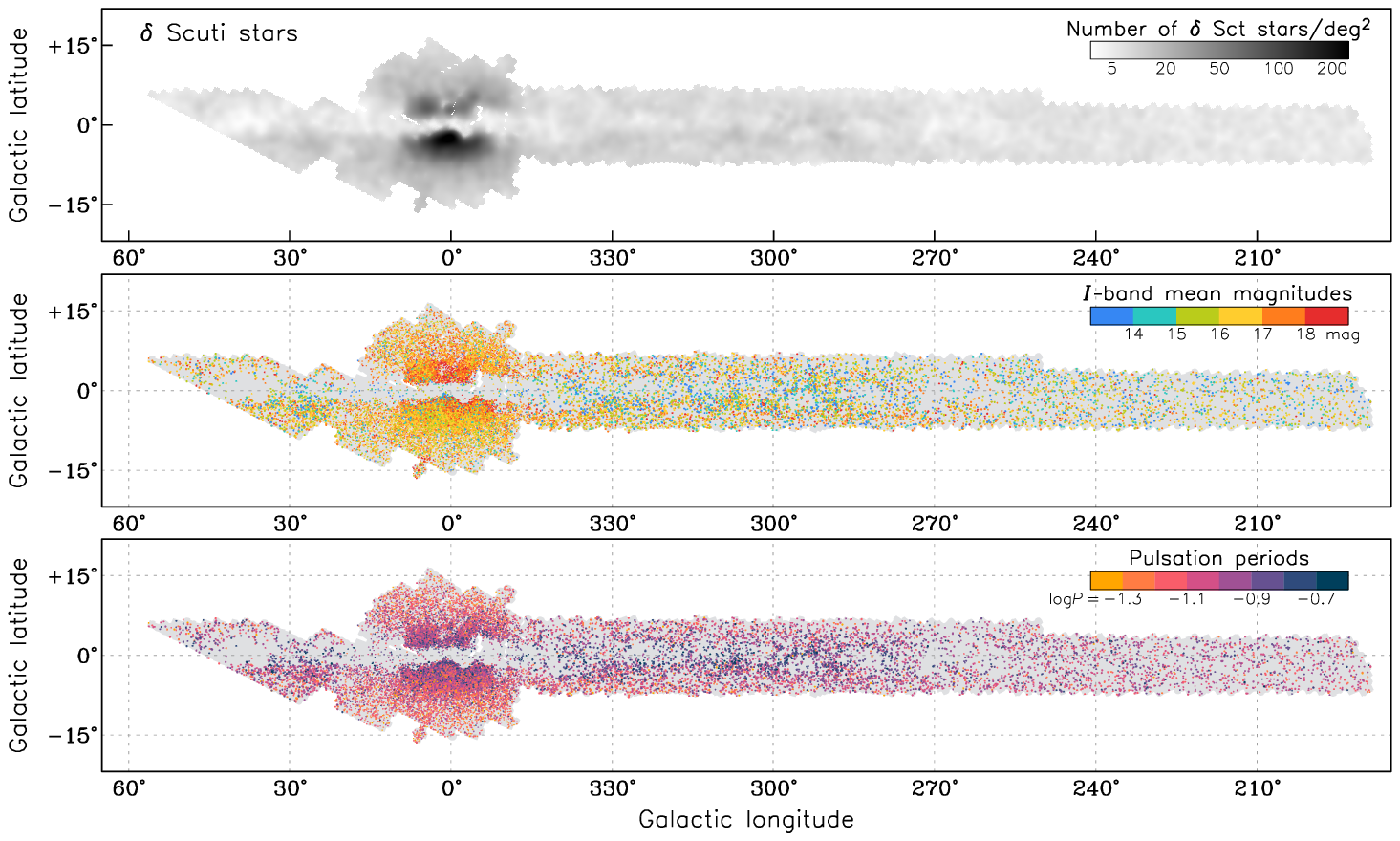}}
\vskip9pt

\FigCap{On-sky distributions of $\delta$~Sct stars in the galactic
  coordinates. {\it Upper panel} displays a surface density map in a
  square root scaling. {\it Middle} and {\it lower panels} show
  positions of individual objects with colors of the points coding
  {\it I}-band mean magnitudes and dominant pulsation periods,
  respectively. The gray area shows the OGLE footprint in the Galactic
  bulge and disk.}
\end{figure}
\end{landscape}

\Section{Completeness and Cross-Check with Other Catalogs}
We cross-matched our collection of Galactic $\delta$~Sct stars with
several catalogs of variable stars to check the accuracy of our
classification and the completeness of our sample. In the VSX, ASAS-SN
and ZTF catalogs of variable stars (Watson \etal 2006, Jayasinghe
\etal 2020, Chen \etal 2020) we found a total of 1757 out of 14\,328
newly selected $\delta$~Sct stars. The great majority of these
previously known stars have been classified as $\delta$~Sct variables
(DSCT, HADS, or SXPHE) in the external catalogs. However, several
dozen objects were assigned to other categories, usually contact
binary systems or RR~Lyr stars. We verified these stars by
re-inspecting their OGLE light curves and in some cases we changed our
classification removing these objects from our sample.

On the other hand, the VSX, ASAS-SN, and ZTF catalogs contain over
1500 objects classified as $\delta$~Sct stars that were not included
in the preliminary version of our collection, despite the fact that
their time-series photometry was available in the OGLE database. We
carefully examined these light curves and completed our collection
with 159 additional $\delta$~Sct variables that were overlooked in our
selection procedure. The rest of the candidates for $\delta$~Sct stars
from other surveys turned out to be eclipsing or ellipsoidal binaries,
low-amplitude variables of vague type, or just constant stars. Some of
the low-amplitude variables with sinusoidal light curves may actually
be $\delta$~Sct pulsators, however we decided not to add these stars
to our collection, because their classification was very uncertain.

The completeness of our collection of $\delta$~Sct stars was assessed
based on the objects with two entries in the OGLE database due to
their position in the overlapping parts of adjacent OGLE fields. We
had the potential to double-detect these objects during our selection
procedure, although the final version of the collection contains only
one entry per star -- usually the one with a larger number of data
points. Considering only light curves with 30 or more data points, we
{\it a posteriori} detected 712 stars with double detections in the
OGLE database, so we had a chance to find 1424 counterparts. We
independently identified 1163 of them, which corresponds to the
completeness of about 78\%.

This moderate completeness reflects the difficulties of identifying
$\delta$~Sct stars with small amplitudes and symmetric light
curves. In contrast to Cepheids or RR~Lyr stars, the number of
$\delta$~Sct variables increases with decreasing amplitudes. Indeed,
most of the missed objects belong to the lowest-amplitude
($A(I)<0.1$~mag) pulsators in our collection. We also checked the
completeness of the sample of high-amplitude $\delta$~Sct stars (HADS)
with the {\it I}-band amplitudes larger than 0.3~mag. As expected, the
completeness of the HADS collection was much higher, reaching 90\%. In
this case, the overlooked objects were almost exclusively the weakest
ones, with {\it I}-band mean magnitudes greater than 18~mag.

\Section{Discussion}
\begin{figure}[p]
\includegraphics[bb = 30 50 520 750, clip, width=12.8cm]{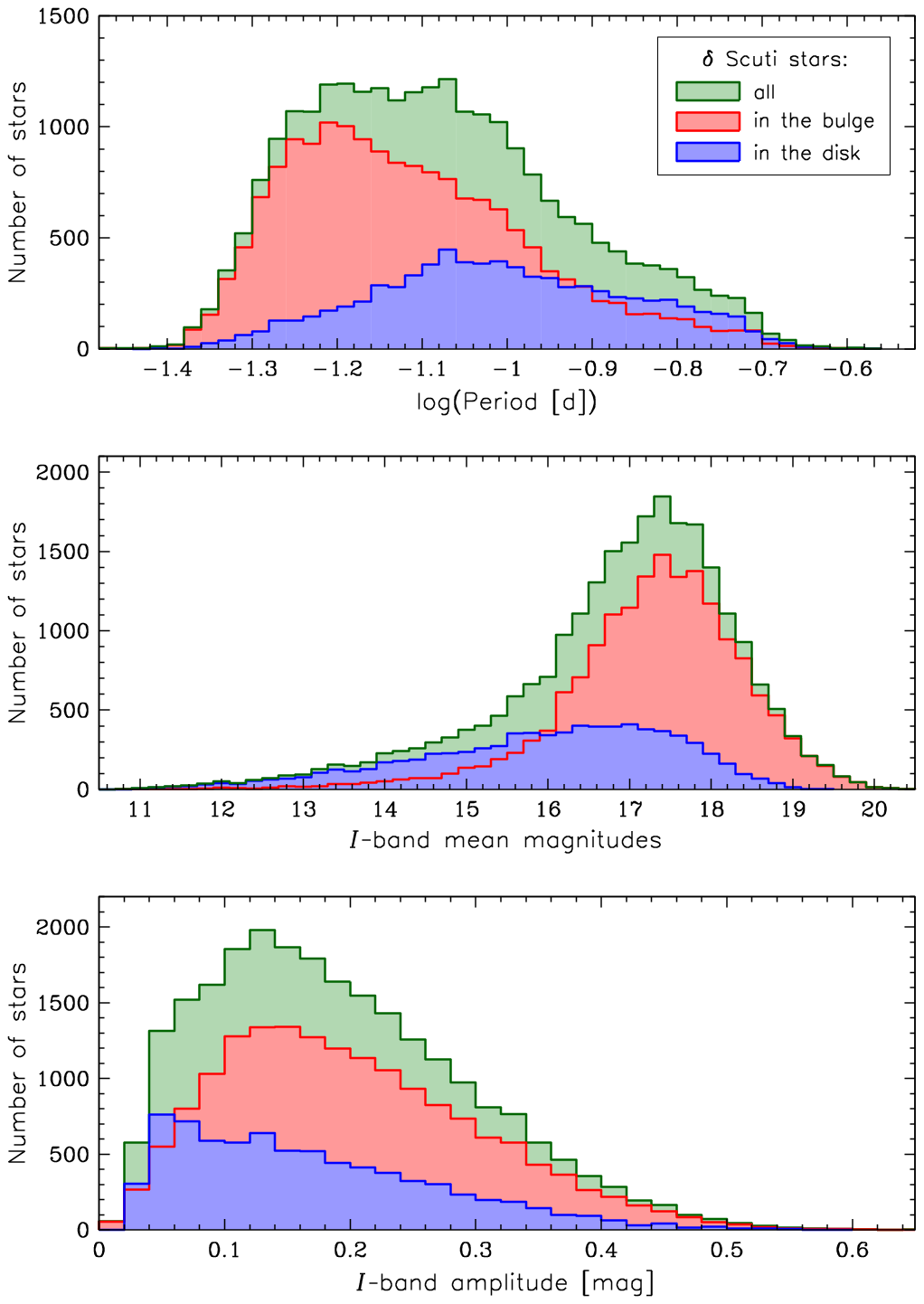}
\FigCap{Distributions of dominant pulsation periods ({\it upper panel}),
{\it I}-band mean magnitudes ({\it middle panel}), and {\it I}-band
peak-to-peak amplitudes ({\it lower panel}) of $\delta$~Sct stars in the
Milky Way. Red histograms show 16\,812 variables identified in the fields
toward the Galactic bulge, blue histograms show distributions of 7676
objects detected in the Galactic disk, green histograms are the sum of the
previous two.}
\end{figure}

\Subsection{Period, Magnitude and Amplitude Distributions}
Fig.~2 displays the distributions of dominant pulsation periods (upper
panel), apparent {\it I}-band mean magnitudes (middle panel), and {\it
  I}-band peak-to-peak amplitudes (lower panel) of 24\,488
$\delta$~Sct stars from our collection. In addition to the total
sample (green histograms), we present the distributions separately for
$\delta$~Sct variables found toward the Galactic bulge (red
histograms) and toward the Galactic disk (blue histograms).

The distribution of pulsation periods of the total sample has a double
maximum: for periods of about 0.063~d ($\log{P}=-1.20$) and 0.085~d
($\log{P}=-1.07$). The former value corresponds to the peak in the
period distribution for the bulge $\delta$~Sct population, while the
latter period coincides with the maximum of the distribution for the
disk variables. These evidently different period distributions of
$\delta$~Sct stars in the bulge and disk must be related to their
different evolutionary history. The shorter-period metal-poor SX~Phe
stars seem to be much more common in the central parts of the Milky
Way, while Population~I $\delta$~Sct stars dominate in the Galactic
disk.

The bulge and disk members also have different luminosity
distributions (middle panel of Fig.~2). In this case, it can be
explained by the different depths of the regular and GVS OGLE
surveys. The bulge sample is largely composed of the $\delta$~Sct
stars detected by Pietrukowicz \etal (2020) in the regular OGLE
fields, where the faintest objects reach luminosities of
$I>20$~mag. The disk variables were mostly identified in the shallower
GVS dataset, so the faintest stars in this sample barely exceed
$I=19$~mag. On the other hand, the saturation limit is respectively
brighter in the GVS photometry, up to $I=11$~mag.

The amplitude distribution (lower panel of Fig.~2) peaks at
$A(I)\approx0.12$~mag. It can be assumed that the completeness of our
collection decreases toward smaller amplitudes and completely
collapses for amplitudes below 0.04~mag. It is interesting that the
amplitude distribution for the disk sample of $\delta$~Sct stars has a
maximum at $A(I)\approx0.05$~mag, which probably indicates that there is
a potential to increase the completeness of our bulge sample.

\Subsection{Multimode Variables}
\begin{figure}[b]
\includegraphics[bb = 30 230 570 750, clip, width=12.5cm]{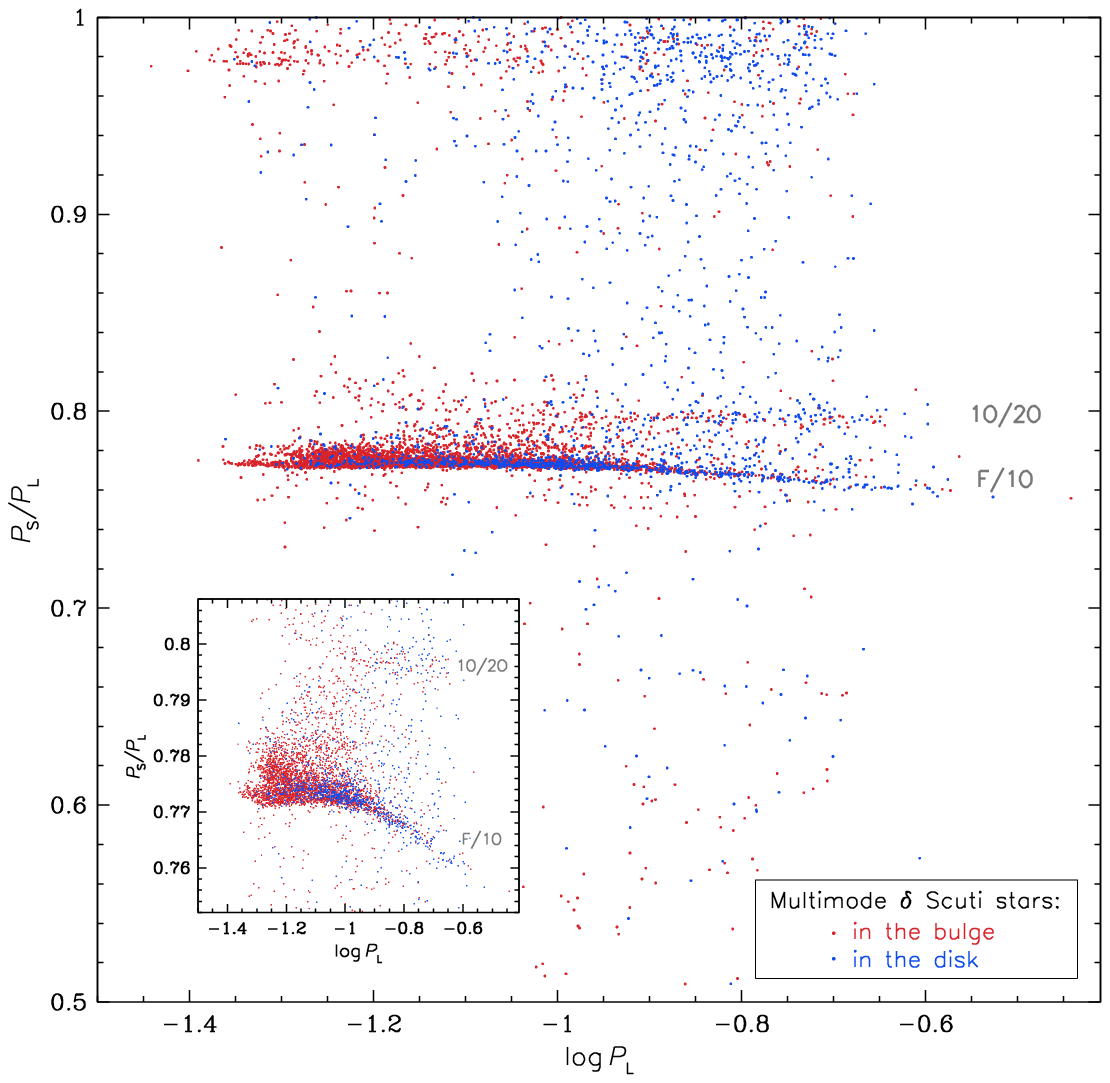}
\FigCap{Petersen diagram for multimode $\delta$~Sct stars in the Galactic
bulge (red points) and disk (blue points). The inset shows a zoom-in
to the sequences formed by F/1O and 1O/2O variables.}
\end{figure}

Multiperiodic $\delta$~Sct stars are useful in asteroseismological
studies, because two or more accurately measured pulsation modes yield
strong constraints on the stellar mass, luminosity, and chemical
composition. Recently, Netzel \etal (2021) studied the OGLE
collection of 10\,092 $\delta$~Sct stars in the Galactic bulge
(Pietrukowicz \etal 2020) and identified 3083 objects pulsating in
two, three or even four radial modes simultaneously. The
characteristic period ratios of the radial modes were used for their
identification. Additionally, a large fraction of $\delta$~Sct
variables exhibit nonradial pulsation modes (\eg Pietrukowicz \etal
2015), which, however, are far more difficult to identify, because
they may form a variety of period ratios with other modes.

We performed a search for multiperiodic $\delta$~Sct stars using the
standard method. From each {\it I}-band light curve, the dominant
frequency and its harmonics were subtracted by fitting a truncated
Fourier series.  Then, the period search was conducted on the residual
data and the process of the light curve prewhitening and frequency
analysis was repeated. Finally, we visually inspected the light curves
with the largest signal-to-noise ratios of the secondary and tertiary
periods. In this way, we selected only the stars with the most
distinct additional periodicities.  Thus, the completeness of our
sample of multimode $\delta$~Sct pulsators can be improved by
including periodicity associated with smaller amplitudes, but at the
cost of lower purity of the sample.

Fig.~3 shows the Petersen diagram for the detected multimode
$\delta$~Sct stars. Red and blue points indicate objects found in the
Galactic bulge and disk regions, respectively. The bulge variables
have on average shorter periods, but generally both populations follow
the same sequences in the Petersen diagram. The most conspicuous
structure in this diagram is the sequence for period ratios ranging
from $\approx0.76$ to $\approx0.78$ formed by the fundamental (F) and
first-overtone (1O) pulsation modes. The second sequence at
$\approx0.80$ period ratios corresponds to the first and second (2O)
overtones. Both sequences are zoomed in the inset in Fig.~3 to reveal
details of these structures. Note, for example, that the F/1O sequence
splits into two branches in the short-period end: for periods ratios
of about 0.773 and 0.778.

\vskip3pt
Approximately a third of the selected multiperiodic $\delta$~Sct stars
are located outside the F/1O and 1O/2O sequences in the Petersen
diagram. This group includes nonradial pulsators as well as double-,
triple-, or even quadruple-mode pulsators (Netzel \etal 2021) with
various combinations of the radial modes excited. The former category
includes stars with a close pair of frequencies, which are visible as
a surplus of points at the top of the Petersen diagram
($P_\mathrm{S}/P_\mathrm{L}\gtrsim0.95$). The most plausible
explanation for these close doublets of frequencies is the excitation
of nonradial modes by resonance with the main period of oscillation.

\vskip3pt
The identification of the pulsation modes is not a simple task for
many $\delta$~Sct variables, therefore we do not include this
information in our collection. We just divide our sample into
singlemode and multimode pulsators and provide up to three dominant
periods for each star.

\Subsection{$\delta$~Sct Stars in Binary Systems}
\vskip3pt
Binary systems with oscillating components are powerful tools for
probing the absolute stellar parameters of the pulsating stars:
masses, radii, and luminosities. Known binaries containing
$\delta$~Sct stars are much more common than systems containing
classical Cepheids (\eg Udalski \etal 2015b, Pilecki \etal 2021) or
RR~Lyr stars (\eg Hajdu \etal 2021). Kahraman Ali{\c{c}}avu{\c{s}}
\etal (2017) published a list of 92 eclipsing binary systems with a
$\delta$~Sct component, while Liakos and Niarchos (2017) cataloged as
many as 199 eclipsing, ellipsoidal, visual, and spectroscopic binaries
containing $\delta$~Sct stars.

\begin{figure}[p]
\includegraphics[bb = 35 50 530 760, clip, width=12.8cm]{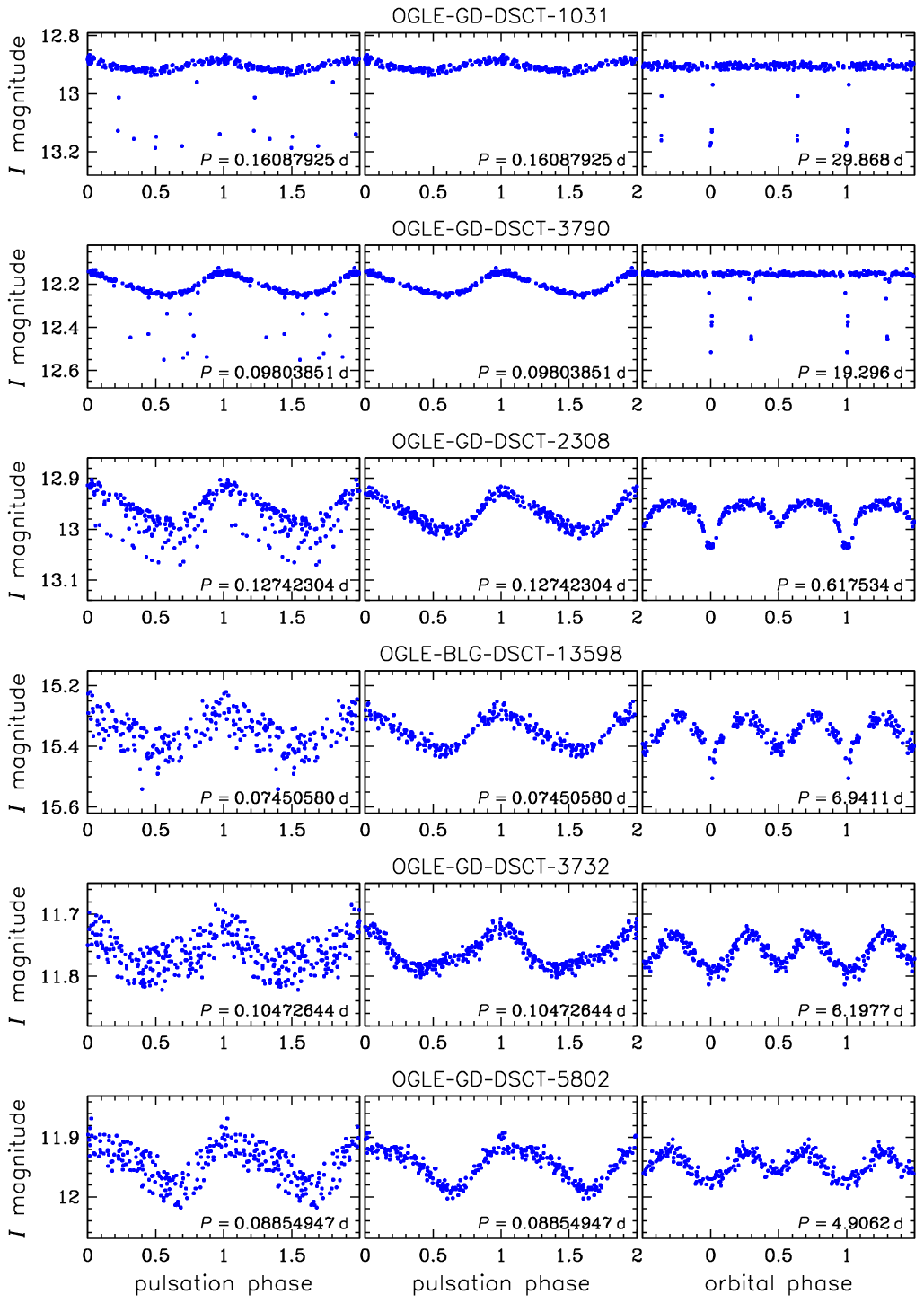}
\FigCap{Example {\it I}-band light curves of $\delta$~Sct stars with
additional eclipsing (four {\it upper panels}) or ellipsoidal (two {\it
lower panels}) modulation. {\it Left panels} show original light curves
folded with the pulsation periods. {\it Middle} and {\it right panels}
present pulsation and eclipsing/ellipsoidal light curves, respectively,
after subtracting the other component.}
\end{figure}

\vskip3pt
Binary systems with $\delta$~Sct components are also present in our
collection. Pietru\-kowicz \etal (2020) listed 14~$\delta$~Sct
variables with additional eclipsing or ellipsoidal modulation. In this
paper, we expand this list with additional 20~systems. We divided this
sample into detached systems, semidetached systems and ellipsoidal
variables. Table~1 provides the most important parameters of these
objects, while Fig.~4 shows light curves of six objects from this
sample.
\MakeTableee{l@{\hspace{6pt}}
c@{\hspace{4pt}}   
c@{\hspace{6pt}}
c@{\hspace{8pt}}
c@{\hspace{4pt}}
l@{\hspace{0pt}}
c@{\hspace{0pt}}}
{12.5cm}{New $\delta$~Sct stars with additional eclipsing or ellipsoidal
modulation}
{\hline
\noalign{\vskip3pt}
\multicolumn{1}{c}{Identifier}
& R.A.
& Dec.
& $\langle{I}\rangle$
& $P_{\rm puls}$
& \multicolumn{1}{c}{$P_{\rm orb}$}
& Binarity \\
& [J2000.0]
& [J2000.0]
& [mag]
& [d]
& \multicolumn{1}{c}{[d]}
& type \\
\noalign{\vskip3pt}
\hline
\noalign{\vskip3pt}
OGLE-GD-DSCT-0393   & 06\uph34\upm28\zdot\ups31 & $+09\arcd23\arcm38\zdot\arcs2$ & 13.755 & 0.06005839 & ~~4.9618 & EA \\
OGLE-GD-DSCT-1031   & 07\uph31\upm39\zdot\ups71 & $-28\arcd45\arcm05\zdot\arcs9$ & 12.919 & 0.16087925 & 29.868  & EA \\
OGLE-GD-DSCT-1398   & 08\uph00\upm23\zdot\ups53 & $-23\arcd35\arcm35\zdot\arcs5$ & 14.081 & 0.14320227 & ~~5.0375 & EA \\
OGLE-GD-DSCT-1423   & 08\uph03\upm11\zdot\ups27 & $-36\arcd47\arcm52\zdot\arcs9$ & 14.560 & 0.11235700 & 18.567  & EA \\
OGLE-GD-DSCT-2308   & 09\uph45\upm56\zdot\ups82 & $-59\arcd50\arcm41\zdot\arcs9$ & 12.968 & 0.12742304 & ~~0.617534 & EB \\
OGLE-GD-DSCT-3010   & 11\uph04\upm00\zdot\ups73 & $-56\arcd39\arcm04\zdot\arcs6$ & 13.050 & 0.08425622 & ~~7.4180 & Ell \\
OGLE-GD-DSCT-3494   & 12\uph05\upm49\zdot\ups30 & $-58\arcd32\arcm49\zdot\arcs4$ & 15.390 & 0.08109499 & ~~6.2934 & Ell \\
OGLE-GD-DSCT-3732   & 12\uph36\upm23\zdot\ups93 & $-58\arcd58\arcm20\zdot\arcs2$ & 11.760 & 0.10472644 & ~~6.1977 & Ell \\
OGLE-GD-DSCT-3790   & 12\uph44\upm18\zdot\ups26 & $-66\arcd40\arcm29\zdot\arcs0$ & 12.224 & 0.09803851 & 19.296  & EA \\
OGLE-GD-DSCT-5612   & 16\uph08\upm36\zdot\ups57 & $-55\arcd51\arcm11\zdot\arcs2$ & 14.310 & 0.07350542 & ~~3.1868 & EB \\
OGLE-GD-DSCT-5802   & 16\uph20\upm54\zdot\ups45 & $-54\arcd55\arcm19\zdot\arcs4$ & 11.946 & 0.08854947 & ~~4.9062 & Ell \\
OGLE-GD-DSCT-5901   & 16\uph27\upm08\zdot\ups32 & $-40\arcd46\arcm36\zdot\arcs6$ & 16.681 & 0.08709063 & ~~3.4373 & EB \\
OGLE-GD-DSCT-5979   & 16\uph32\upm26\zdot\ups61 & $-53\arcd11\arcm11\zdot\arcs2$ & 15.758 & 0.07376154 & ~~6.6009 & Ell \\
OGLE-GD-DSCT-6092   & 16\uph39\upm21\zdot\ups85 & $-45\arcd46\arcm26\zdot\arcs7$ & 14.530 & 0.11055747 & ~~8.8291 & Ell \\
OGLE-GD-DSCT-6454   & 17\uph12\upm13\zdot\ups12 & $-50\arcd29\arcm24\zdot\arcs5$ & 14.299 & 0.08480399 & ~~5.5759 & Ell \\
OGLE-GD-DSCT-6772   & 18\uph43\upm11\zdot\ups85 & $-08\arcd45\arcm22\zdot\arcs6$ & 15.734 & 0.07839580 & ~~3.6596 & Ell \\
OGLE-BLG-DSCT-11682 & 17\uph15\upm33\zdot\ups49 & $-17\arcd06\arcm37\zdot\arcs3$ & 16.344 & 0.05725150 & ~~2.8922 & Ell \\
OGLE-BLG-DSCT-13598 & 17\uph55\upm27\zdot\ups56 & $-37\arcd54\arcm38\zdot\arcs6$ & 15.352 & 0.07450581 & ~~6.9411 & EB \\
OGLE-BLG-DSCT-14775 & 18\uph23\upm04\zdot\ups05 & $-29\arcd21\arcm24\zdot\arcs0$ & 15.967 & 0.05889860 & ~~3.7133 & Ell \\
OGLE-BLG-DSCT-16533 & 18\uph49\upm50\zdot\ups75 & $-16\arcd46\arcm06\zdot\arcs7$ & 13.889 & 0.08021462 & ~~5.1894 & Ell \\
\noalign{\vskip3pt}
\hline}

\begin{figure}[t]
\includegraphics[bb = 35 170 530 760, clip, width=12.8cm]{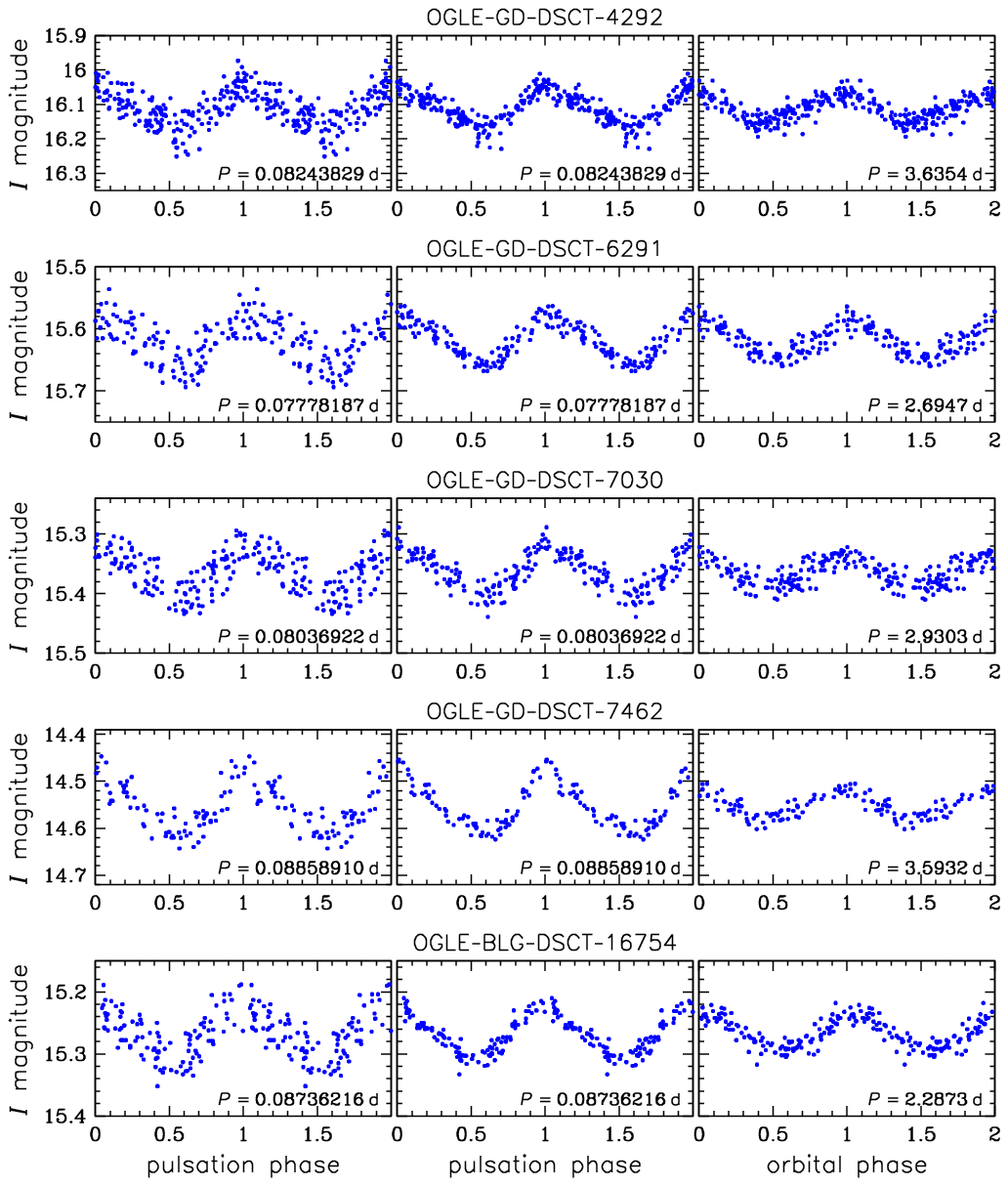}
\FigCap{{\it I}-band light curves of $\delta$~Sct stars with additional
long-period modulation. {\it Left panels} show original light curves folded
with the pulsation periods. {\it Middle} and {\it right panels} show
disentangled light curves folded with the pulsation and long period,
respectively.}
\end{figure}
\MakeTable{l@{\hspace{6pt}}
c@{\hspace{6pt}}   
c@{\hspace{8pt}}
c@{\hspace{8pt}}
c@{\hspace{8pt}}
c@{\hspace{3pt}}}
{12.5cm}{$\delta$~Sct stars with additional long-term modulation}
{\hline
\noalign{\vskip3pt}
\multicolumn{1}{c}{Identifier}
& R.A.
& Dec.
& $\langle{I}\rangle$
& $P_{\rm puls}$
& $P_{\rm long}$ \\
& [J2000.0]
& [J2000.0]
& [mag]
& [d]
& [d] \\
\noalign{\vskip3pt}
\hline
\noalign{\vskip3pt}
OGLE-GD-DSCT-4292   & 13\uph47\upm34\zdot\ups65 & $-64\arcd09\arcm00\zdot\arcs7$ & 16.111 & 0.08243829 & 3.6354 \\
OGLE-GD-DSCT-6291   & 16\uph52\upm20\zdot\ups90 & $-50\arcd01\arcm54\zdot\arcs3$ & 15.619 & 0.07778187 & 2.6947 \\
OGLE-GD-DSCT-7030   & 18\uph52\upm58\zdot\ups95 & $-07\arcd34\arcm13\zdot\arcs9$ & 15.363 & 0.08036922 & 2.9303 \\
OGLE-GD-DSCT-7462   & 19\uph07\upm01\zdot\ups75 & $+14\arcd02\arcm37\zdot\arcs6$ & 14.549 & 0.08858910 & 3.5932 \\
OGLE-BLG-DSCT-16754 & 19\uph00\upm30\zdot\ups22 & $-15\arcd08\arcm30\zdot\arcs0$ & 15.266 & 0.08736216 & 2.2873 \\
\noalign{\vskip3pt}
\hline}

\vskip3pt 
In addition, we found five $\delta$~Sct stars with well pronounced
secondary periods in the range 2--4~d, for which we did not detect
alternating shallower and deeper minima characteristic for ellipsoidal
variables (Fig.~5, Table~2). We do not rule out that these long-period
modulations are caused by binarity (in such case the orbital periods
would be twice as long as the measured ones), but the changes might
also be caused by rotation of spotted stars or by nonradial
oscillations as $\gamma$~Doradus stars. Such hybrid
$\delta$~Sct/$\gamma$~Dor variables are well known (\eg Henry and
Fekel 2005) and there is no wonder that they could also appear in our
collection.

\Subsection{Globular Clusters Members}
Among over 10\,000 $\delta$~Sct variables detected toward the Galactic
bulge, Pietru\-kowicz \etal (2020) distinguished 22 objects that are
located in the sky not farther than three half-light radii ($3r_h$)
from the centers of nine globular clusters. Most of these stars are
probably SX~Phe variables that are cluster members, but some of them
may be field objects accidentally located along the same line-of-sight
as the cluster.
\MakeTable{l@{\hspace{8pt}}
c@{\hspace{8pt}}   
c@{\hspace{8pt}}
c@{\hspace{8pt}}
c@{\hspace{8pt}}
c@{\hspace{3pt}}}
{12.5cm}{$\delta$~Sct stars located within the radius of $3r_h$ from centers of globular clusters}
{\hline
\noalign{\vskip3pt}
\multicolumn{1}{c}{Cluster}
& \multicolumn{1}{c}{Identifier}
& $\langle{I}\rangle$
& $P_{\rm puls}$
& $r/r_h$
& \multicolumn{1}{c}{Other name} \\
\noalign{\vskip3pt}
\hline
\noalign{\vskip3pt}
M62      & OGLE-BLG-DSCT-10730 & 16.357 & 0.09267790 & 0.4 & \\
M19      & OGLE-BLG-DSCT-10810 & 17.536 & 0.06571908 & 1.1 & \\
NGC 6287 & OGLE-BLG-DSCT-10961 & 17.835 & 0.06571120 & 0.5 & \\
Ton 2    & OGLE-BLG-DSCT-12743 & 17.767 & 0.14461103 & 2.6 & \\
         & OGLE-BLG-DSCT-12754 & 18.368 & 0.11930232 & 1.8 & \\
NGC 6441 & OGLE-BLG-DSCT-13369 & 16.702 & 0.08992963 & 2.6 & V1517 Sco \\
NGC 6541 & OGLE-BLG-DSCT-14113 & 16.547 & 0.05746375 & 1.5 & \\
         & OGLE-BLG-DSCT-14119 & 16.149 & 0.06466907 & 0.1 & V0825 CrA \\
         & OGLE-BLG-DSCT-14120 & 16.180 & 0.06571376 & 0.3 & V0826 CrA \\
NGC 6558 & OGLE-BLG-DSCT-14210 & 16.777 & 0.06851215 & 1.4 & \\
M28      & OGLE-BLG-DSCT-14983 & 17.234 & 0.06551779 & 2.8 & \\
NGC 6638 & OGLE-BLG-DSCT-15476 & 16.825 & 0.05766997 & 2.2 & \\
\noalign{\vskip3pt}
\hline}

In this work, we extend this list by additional 12~$\delta$~Sct
(SX~Phe) stars positionally coincident with nine globular clusters. In
Table~3, we present these variables found in the regions outlined by
three half-light radii of the clusters. We used this condition to
comply with the work by Pietrukowicz \etal (2020), however different
criteria would result in significantly different numbers of candidate
cluster members. For example, as many as 71 $\delta$~Sct variables are
located within one tidal radius from the center of a globular
cluster. Future investigations, in particular analysis of their proper
motions, should answer the question of how many of these stars are
physically associated with the clusters.

\Section{Conclusions}
We release the largest sample of $\delta$~Sct stars in the Milky
Way. Our collection includes about 10\,000 variables identified by
Pietrukowicz \etal (2020) in the OGLE Galactic bulge fields and over
14\,000 new objects detected in the GVS fields in the Galactic disk
and outer bulge.  The vast majority ($\sim$88\%) of these new
$\delta$~Sct stars have not been reported by other sky surveys so
far. For each object, we provide its long-term OGLE light curves
spanning over 20 years (OGLE-II + OGLE-III + OGLE-IV) in some
cases. These data are available through a user-friendly WWW
interface. Our collection provides a framework for better
understanding of pulsations and evolution of intermediate-mass stars.

\Acknow{This work has been supported by the National Science Centre,
Poland, grant MAESTRO no. 2016/22/A/ST9/00009. MG is supported by the
EU Horizon 2020 research and innovation programme under grant agreement No
101004719. This research has made use of the International Variable Star
Index (VSX) database, operated at AAVSO, Cambridge, Massachusetts, USA.}

\end{document}